\documentclass{article}
\pdfoutput=1

\usepackage{lscape}
\usepackage{arxiv}
\usepackage{amsmath,amssymb,amsfonts,latexsym}
\usepackage{graphicx,psfrag}
\usepackage{subcaption}
\makeatletter
\def\endthebibliography{%
	\def\@noitemerr{\@latex@warning{Empty `thebibliography' environment}}%
	\endlist
}
\makeatother

\begin{document}

\title{Learning from Others in the Financial Market}

\author{Matthias Feiler, Thibaut Ajdler
\thanks{mail: matthias.feiler@lgt.com, thibaut ajdler@lgt.com}}

\maketitle

\begin{abstract}
Prediction problems in finance go beyond estimating the unknown parameters of a model (e.g of expected returns). This is because such a model would have to include parameters governing the market participants' propensity to change their opinions on the validity of that model. This leads to a well--known circular situation characteristic of financial markets, where participants collectively create the future they wish to estimate. In this paper, we introduce a framework for organizing multiple expectation models and study the conditions under which they are adopted by a majority of market participants.  
\end{abstract}

\section{Introduction}

When applying machine learning to the natural or the social sciences the difference is in the level at which patterns are identified and exploited: in natural science, data may be reproduced through repeated probing or experiments, which leads to its classification as an ``exact'' science. In social science, we study the possibly infinite ways in which people behave given the data. As an example, a natural scientist might be interested in estimating the fill level of oil reservoirs using satellite images, while a social scientist, in particular in the field of finance, worries about how the knowledge about fill levels affects investors' estimates of the scarcity of the commodity. In the latter case, it cannot be assumed that fill levels alone determine the trading levels of oil but other events, e.g. geopolitics may at the same time impact the future supply of oil. The real question is what we believe the majority of other market participants will regard as the \textit{relevant} piece of information. The financial market is a beauty contest in which majorities decide on the winner, and objective criteria are only the contestants in a competition for the attention of investors \cite{keynes2018general}. 

Attempts have been made in the economic literature to get around the complexity arising from the self--referential nature of financial markets. An influential concept was the hypothesis put forward by Muth \cite{muth1961rational} that the agents' predictions of future payoffs are not systematically wrong: agents are capable of forming rational expectations which take all available information into account. Grossmann \cite{grossman1976efficiency} proposes the idea that prices aggregate information that is initially dispersed across investors. The assumption is that the \textit{true} (fundamental) price is known to some but not all market participants leading to an information asymmetry. The asymmetry is reduced when participants observe prices which convey information about the aggregate demand and, from there, the true price of a traded asset. Even though this inference may be too complicated to be carried out by the average market participant, market forces act \textit{as if} this was the average behavior as traders with incorrect estimates would incur losses and would eventually be driven out of the market. Only traders with correct beliefs survive. This is referred to as the market selection argument for the information efficiency of prices \cite{blume2006if}. 

The general viewpoint adopted by the efficient market hypothesis is that the ``true'' (fundamental) value of an asset is somehow exogenously given and remains unchanged during the process of price discovery. Efficiency means that traded prices reveal everything there is to be known about the security. Even if we accept the idea of a true value we find that in practice, i.e. over finite investment horizons, \textit{value} is relative as the price obtained when selling a security at some future instant of time will be set by the aggregate demand at that instant. This leads to the problem of having to determine what others will be ready to pay at the end of our investment period. 

A different viewpoint is the following: Fundamental value cannot exist \textit{ex ante} if it depends on the choices of the security market itself. The payoff of assets to investors depends on future states of the economy. All investors face the same problem of not knowing these states and, more importantly, the impact of these states on the decisions of other investors. As a result, estimates about the future are ``irreducibly subjective'' \cite{orlean201214}, they are no more than opinions. ``Investors have to position themselves relative to the market opinion and by doing so they bring it to life'' [ibid.]. Fundamental value is co--constructed in a framework where multiple assessments about future payoffs compete to become the dominant way of interpreting current events.

When agents factor in the presence of other agents who also try to forecast the value of the asset we potentially run into an infinite regress problem \cite{townsend1983forecasting}. Allen, Morris and Shin \cite{allen2006beauty} study the problem in a homogeneous market populated by Bayesian rational agents and find that there is a bias towards prior information, when agents forecast the forecasts of others. In a recent dicussion paper, Beckert argues that decision--making is instead anchored in imagined, ``fictional'' depictions of future outcomes which may include genuine novelty. Such ``imagineries can help fashion shared visions and social narratives that coordinate future--oriented beliefs'' \cite{beckert2013imagined}. But if the value of an asset depends on what the market as a whole thinks and the thinking, in turn, depends on its un--observable imaginative capacity, is financial forecasting possible at all?

Practical experience indicates that investors' expectations are guided by conventions, i.e. generally accepted ideas or leading paradigms in which the market operates. The aim of this paper is to introduce a framework for identifying the prevailing market convention. This may be seen as the narrative adopted by a majority of investors when forming expectations about future asset payoffs. For example, an inverted US yield curve (long--term bond yields $<$ short--term yields) may be seen as a sign of an incumbent recession (narrative 1) or as a consequence of global demand for US bonds in a low--yield world (narrative 2). To decide on one of the alternatives, we observe the statistical dependencies of asset returns on available signals where each ``signal'' represents a simple forecasting model. The key intuition in our approach is that, over some interval of time, the same model may be good for multiple different asset classes (i.e. it is a ``factor'' behind the asset returns). This gives rise to a hierarchy of models which enables us to determine which forecast is closest to the majority opinion. 

\section{Problem Statement}

Suppose the uncertainty around a price target $p^*$ can be described by two scenarios $p_1^*$ and $p_2^*$. The question is which of the two fair values is likely to come ``true'' over future trading sessions. If we knew all agents who support $p_1^*$ or $p_2^*$ (as well as their susceptibility to change opinions) we could base our decision on the current relative importance of the scenarios. If market participants are able to change their opinion (which they are) the effective ``true'' price of the asset becomes a moving target. A standard way then would be to observe outcomes, i.e. traded prices $p$ over time. This would eventually reveal whether $p$ was generated from a distribution centered around $p_1^*$ or $p_2^*$. However, this is a slow process that relies on the assumption of an even lower switching rate between the two alternatives. 

Our approach is to decide on the basis of the performance of related assets, i.e. to collect contemporaneous evidence in a cross--section of returns. The rationale is that opinion formation around price targets is made for all investable assets simultaneously using similar, overlapping arguments. We define the relation among assets indirectly using common links they may have to the available signals. If multiple assets depend on similar signals they form a group. As new data becomes available the signal dependencies are re--evaluated. If the group persists we conclude that the common signal is still a valid one. If, on the other hand, the group members connect to an entirely new set of signals we suspect that a regime shift has taken place and the dependencies --even of assets that still connect to the original signal-- may need to be updated. The signal sensitivities differ among assets and give rise to different speeds at which the information contained in the signal is priced in. Within any group there will be members that are more susceptible to a change in driving factors. The idea is to systematically use this information in order to select relevant signals for the investment decision at hand. 

\subsection*{Formal Description}

Let $x$ be an $m$--dimensional vector of asset returns and $s$ be an $n$--dimensional vector of signals. Returns are obtained at instant of time $t+1$ due to investments made at instant $t$. We assume that investors hold their positions over a period $[t,\ t+Q-1]$, $Q>1$. During that period they collect signals $s_t$ to prepare for the following trading decision at $t+Q$. For simplicity we assume that asset returns are binary variables $x \in \{0, \, 1\}^m$ where $1$ corresponds to $x>\bar x$ above some threshold $\bar x > 0$. Likewise $s \in \{0, \, 1\}^n$, which allows us to combine both continuous inputs $s>\bar s$, $\bar s > 0$, as well as discrete occurrences (e.g. of news articles). In summary, our market model consists of the multinomial distribution $p(X_{t+1}, S_t)$ where the decisions (based on $S_t$) and outcomes $X_{t+1}$ are separated by one discrete time--step. Notice that we use the subscripts generically to denote any time instant $t$ and its follower $t+1$. $p(X_{t+1}, S_t)$ will be estimated over multiple realizations $\{x_{\tau+1},s_{\tau}\}$, $\tau \in [t,\ t+Q-1]$. The way signals are connected to outcomes depends on a common understanding of the relevance of the information contained in the available signals. This understanding may vary over time. 

As an example, a weak currency is in general good for the local equity market of a country as it makes the export sector more competitive. However, the weakness may be an indication of political uncertainty in which case it must not be used as a signal to invest. Depending on the relative importance assigned to the arguments the new price target is $p_2^*$ or $p_1^*$. As discussed in more detail below, we approximate $p(X_{t+1}, S_t)$ by a tree. We factorize the joint distribution using a second--order approximation \cite{chow1968approximating} with the additional constraint that target variables appear as leaves in the tree. This is realized by not admitting edges that link target variables back to signals during the greedy search for the minimum spanning tree.  

Let $\mathcal{S}$ denote the subtree representing the distribution of signals $p(S_t)$ (with $X_{t+1}$ marginalized out). We assume that the co--dependence among signals is stationary, i.e. the structure of the tree and its edge weights are constant. Target variables connect to $\mathcal{S}$ via edges corresponding to conditional probabilities $p(X_{t+1}| S_t)$. Let $\mathcal{X}$ be the set of leaf nodes. The problem addressed in this paper is that (while $\mathcal{S}$ is constant) the connection of $\mathcal{X}$ to $\mathcal{S}$ may change over time. We write $\mathcal{T}_t = c_t(\mathcal{X}, \mathcal{S})$ for the overall tree resulting from the connection of signal nodes and leaves at instant $t$. Note that observations up to $(x_t, \, s_{t-1})$ are used in the estimation of $c_t(\cdot,\cdot)$. 

\subsection*{Node $x_i$ and its peers}
  
Suppose we are interested in the return of asset $i$ $X_{i,t+1}$. In the above tree this corresponds to a leaf node $x_i$. The leaf distance given $\mathcal{T}_t$ is defined as an $m$-by-$m$ matrix $D_t(\mathcal{X}) = (d(t)_{ij})$ in which $d(t)_{ij}$ is the sum of edge weights in the shortest path from leave node $x_i$ to $x_j$. We define the neighborhood $N_t(x_i) = \{x_j \,|\,d(t)_{ij} < \theta_i\}$ as the set of nodes within a distance $\theta_i >0$ of $x_i$. The point to note is that there are no direct connections among leaf nodes but there will always be (at least) one signal node $s_l$ on the path between to leaves. These intermediary nodes change as $\mathcal{X}$ re--wires to $\mathcal{S}$. This means also that leaf nodes enter and exit the neighborhood $N_t(x_i)$ as new intermediaries appear which affect the weights on the shortest path between leaves. We define the long--term neighborhood as $N_h(x_i) = \{x_j \,|\sum_{t \in h} \,d(t)_{ij} < \theta_i\}$ where $h = [t-H+1,\, t]$ and $H$ is a long lookback horizon. In other words, $N_h(x_i)$ contains the nodes, that are (statistically) close to $x_i$ over many tree generations. 

The principal idea in this paper is the following: While the general tree $\mathcal{T}_t$ is constructed by maximizing the likelihood of observations (or equivalently minimizing the sum of the edge weights between nodes) the connection of node $x_i$ is such that its distance to $N_h(x_i)$ is minimized. The intuition behind this criterion is that it is reasonable to assume that the distance among peers tends to revert to its long--term average as this corresponds to an average over many different market phases and is therefore independent of the current dependencies (of returns on signals) given by $\mathcal{T}_t$. We claim that the best choice for $x_i$ is to connect to a signal node from which peers appear close according to $D_t(\mathcal{X})$. Other than betting on mean--reversion (of peer distance) this also incorporates information about the signal choices of a relevant sub--segment of the market as the objective is to stay close to \textit{all} peers. We argue that this provides a way for estimating the expectation models of others. We summarize:  

\textit{Problem statement}: Given a tree--shaped signal network $\mathcal{S}$ and a map $c_t(\mathcal{X}, \mathcal{S})$ in which all connections are determined except the one from $x_i$ to $\mathcal{S}$, determine the missing connection such that the distance from $x_i$ to $N_h(x_i)$ is minimized.

\section{A tree representation of dependencies}

In this paper we limit ourselves to a simplified representation of the joint distribution which is based on co--occurrences of ``ones''. Divided by the total number of observations this corresponds to a \textit{hit ratio} of positive returns given signals and is motivated by the fact that agents are more interested in finding profitable trade ideas (signals) than understanding the statistical properties of the market. 

\subsection{Dependencies among signals}

We assume that, prior to our experiment, a large number $H$ of signal realizations are available. From this set, we estimate the signal tree $\mathcal{S}$. Given binary observation vectors $s \in \{0,\, 1\}^n$ we define the co--occurrence matrix
\begin{equation}
\label{eq:CC}
C_h = \bar s^T \bar s 
\end{equation}
where $\bar s$ is the $H \times n$ matrix of observations (with $H$, the number of observations and $n$ the number of signals). Co-occurrences have an obvious interpretation as (negative) distances. In this paper, we define the edge weights as $d_{ij}= -\chi_{ij}^2$ for any pair $(i,\,j)$ of nodes with non--zero entry $\chi_{ij}$ in $C_h$. We compute the minimum spanning tree (MST) as the network path that connects all nodes while minimizing the total edge weight. The total weight corresponds to the maximum sum of \textit{squared} co--occurrences, a quantity that will be seen to relate to the spectrum of $C_h$. We use Prim's algorithm \cite{jarnik1930jistem}, \cite{prim1957shortest} to obtain the adjacency matrix $S_h$ of the MST. In a final step, we order $S_h$ according to its column sums. This puts nodes with a larger number of incoming edges to higher levels in the tree hierarchy.

Trees naturally embed levels of abstraction: parent nodes by definition correspond to events that co--occur with child events which are themselves not directly connected.  A typical situation is given by a set of stocks driven by a common factor plus an idiosyncratic process. The factor is the parent that represents some (but not all) characteristics of the stocks. At an even higher level, the performance of multiple factors may depend on the general macro--environment so nodes such as gdp growth and consumer confidence would qualify as parents to the factor nodes. In summary, as we move towards the root of the tree we expect to see more abstract variables. We wish to make this statement a bit more precise:

\textit{Proposition 1:} If a node in $C_h$ is a parent, it is also closest to the 1st principal eigenvector of the sub-matrix associated with its children.

For the proof, let $s_p$ be a parent node in the MST associated with $C_h$ and let $Q = \bar s_q^T \bar s_q$ be the sub--matrix associated with its children. $\bar s_q$ is the $H \times m$ matrix of observations over $m$ children nodes. Choose $x \in \mathbb{R}^{m}$. Then $\bar s_x = \bar s_q x$ is a real--valued linear combination of binary occurrences in the children nodes. Its co--occurrence with $\bar s_q$ is captured by the vector $q_x = \bar s_q^T \bar s_x = Q x \in \mathbb{R}^m$. We will make use of the $l_2$ norm of $Q$ defined as
\begin{equation}
\label{eq:norm}
\|Q\|_2 = \max_{\|x\|_2 = 1} \|q_x\|_2
\end{equation}

\textit{Proof of proposition 1:} By definition, $Q$ is non--negative and symmetric. This means that there exists a vector $v\geq 0$ such that $Q\,v = \rho(Q)v$ where $\rho(Q)\geq0$ is the spectral radius of $Q$. It follows that the rhs in equation (\ref{eq:norm}) is maximized when $x \propto v$. The sequence $\bar s_v = \bar s_q v$, in turn, is the first principal component of $\bar s_q$. By construction, the MST selects $s_p$ as a parent iff $\sum_{q} \chi_{pq}^2 = \|\bar s_q^T \bar s_p\|_2^2$ is maximal among the available parent nodes (all nodes of $\mathcal{S}$ except the $m$ children). But this means that $\bar s_p$ maximizes the same convex utility (\ref{eq:norm}) as $\bar s_v$ except on a discrete search space defined by the tree $\mathcal{S}$. $\hfill \Box$

\subsection{Attaching return nodes}

While dependencies among signals are assumed to be stationary, the dependency of target variables on signals is time--varying. Our objective is to detect which subset of signals is most relevant for an asset at a given instant of time. This is motivated by the empirical observation that despite the theoretical relevance of multiple driving factors they are not always simultaneously ``at work''. This is well--recognized to be a consequence of the conscious attention allocation by decision makers who are aware of their limited capacity to process all available information \cite{sims2003implications}. According to the theory of rational inattention, decision--makers allocate optimally while taking the cost of information acquisition into account. In this paper we argue, that the decision which information to process and which to ignore should be guided by our peers' information selection. The optimal attention allocation is the one that corresponds to the choice of the majority of investors as this will ultimately drive the demand/ supply of an asset. Unlike the neo--classical approach which involves solving a dynamic optimal control problem on the part of the investors we propose a simpler criterion which we believe is closer to describing actual market behavior. Our main argument is that if processing capacity is indeed limited it should not be spent on the problem of optimally selecting which information to follow. Our criterion is that investors simply follow the choices of their peers. 

In the present framework we associate with every asset $i$ (a group of) investors trading that asset. At every instant, the investors decide on important success factors for the asset they hold. We assume that this thought process can be reconstructed by measuring the co--occurrence of positive returns $x_{t}$ and preceding ``buy'' signals $s_{t-1}$. Our measurement starts at a particular instant $t$ and includes $Q$ realizations of signal--return pairs up to instant $t$. We combine these measurements in a vector $\xi_t = [x_t\ s_{t-1}] \in \{0,\,1\}^{m+n}$ and obtain the co--occurrence matrix 
\begin{equation}
\label{eq:CCi}
C_t = \bar \xi_t^T \bar \xi_t 
\end{equation}
which is the ``instantaneous'' version of equation (\ref{eq:CC}) where $\bar \xi_t$ is the $Q\times (m+n)$ matrix of observations over the short--run. $m$ is the number of target variables to be attached to the signal tree $S_h$ as follows. 

Let $S_{C_t}$ be the sub--matrix in $C_t$ corresponding to the co--occurrence of signals. We replace $S_{C_t}$ by the stationary adjacency matrix $S_h$ thereby effectively discarding all short--run relations that might appear in the data collected over the last $Q$ instants. By assumption the dependencies in $S_{C_t}$ are stationary so any deviation from $S_h$ has to be regarded as non--informative. By contrast, the way the target variables depend on signals is allowed to change over time and we assume that $Q$ is sufficiently large for the measured co--occurrences to be statistically significant. Every target node except the one of interest $x_i$ is attached to the signal to which it is closest in the short--run. This means that in every row $x_j$, $j \neq i$ the largest co--occurrence entry is chosen while all others are set to zero. Notice that in our construction, there are no direct links among target nodes, which means that the sub--matrix $X_{C_t}$ corresponding to asset returns in $C_t$ is set to zero. Also, for obvious reasons, we do not include any dependencies of signals on future returns in our model, which means that the upper right $m\times n$ block is also set to zero. At this stage, the final tree $T_t$ is almost specified except for the crucial node $x_i$ which is the asset we want to trade. 

\subsection{Connecting the target node $x_i$}

The row associated with $x_i$ contains all short--run co-occurrences of returns $i$ with signals. Instead of attaching $x_i$ to the closest signal (like all the other nodes) we attach $x_i$ to the signal that allows $x_i$ to stay close the all its peers, i.e. the members $x_j$ of $N_t(x_i)$, $j\neq i$. Due to the short--run nature of our data collection, $x_j$ may connect to very different nodes in $S_h$ over successive trading rounds. This partially reflects noise in the data but also potential early signs of a regime change in the sense that a different signal is becoming relevant for predicting $x_j$. Our idea is to exploit the cross--section $N_t(x_i)$ in order to distinguish information from noise. 

As described above, all return nodes in $X_{t+1}$ are connected to each other through the signal tree $S_h$. Connections occur at all levels depending on the short--run co-dependence of returns on signals: some nodes $x_j$ share the same parent node in the signal tree while others have a common ancestor at lower levels of the tree. In either case the problem for $x_i$ is to choose a node in $S_h$ which allows $x_i$ to connect to $N_t(x_i)$ on the shortest path. Before continuing we introduce a useful device which allows us to identify pathways through the tree. 

We define $\bar S_h$ as the transpose of $S_h$ in which all nonzero entries are set to 1 and let $1_j$ be the $m$-dimensional binary vector indicating the location of node $i$ in $X_{t+1}$. We define the recursion 
\begin{equation}
\label{eq:bwo}
z_l = \bar S_h z_{l+1}  \quad z_L = 1_j
\end{equation}
governing the (unique) transitions among tree levels, starting at the leaf level $z_L$. Let $A_j=\{\bar S_h 1_j, \bar S_h^2 1_j,\dots,\bar S_h^k 1_j=z_0\}$ be the ancestors of node $x_j$, i.e. the set of parents visited until the root of the tree $z_0$ is reached after $k$ iterations. We define  
\begin{equation}
\label{eq:uo}
A_t(x_i) = \bigcup_{j}\, A_{j} \quad \mbox{where } j\,|\, x_j \in N_t(x_i)
\end{equation}
the set union of parent nodes visited by members of the neighborhood of $x_i$ at instant $t$. Notice that also $A_j$ depends on time as the initial condition $z_L$ will be different depending on where $x_j$ attaches to $S_h$ given the short--run data $C_t$. The index $t$ is omitted for notational convenience. $A_t(x_i)$ defines a sub-tree of $S_h$ with root $\alpha_t$. Let $A_{t,\alpha}$ be the set of ancestors of $\alpha_t$ within the complete tree $S_h$. We define  
\begin{equation}
\label{eq:uo}
O_t(x_i) = A_t(x_i)\setminus A_{t,\alpha}
\end{equation}
This set corresponds to the sub--tree connecting the members of $N_t(x_i)$ to each other (including its root $\alpha_t$). 

\textit{Proposition 2:} The set $O_t(x_i)$ is an equivalence class with respect to the problem of finding the shortest path between $x_i$ and the elements of the set $N_t(x_i)$. 

\textit{Proof:} Nodes in $N_t(x_i)$ are by construction endpoints of $S_h$. Since $S_h$ is a tree, every edge leading to endpoints has to be visited. This means that the distance to all nodes within the sub--tree $O_t(x_i)$ is equal to the sum of edge weights of $O_t(x_i)$ independent to which of its members $x_i$ connects.$\hfill \Box$

It follows that $O_t(x_i)$ can be regarded as a single node $s^*$ within $S_h$. The distance to $s^*$ is set equal to the minimum weight of the edges from $x_i$ to any member of $O_t(x_i)$. The remaining problem is to find the shortest path from $x_i$ to $s^*$ through the remaining tree of $S_h$ i.e. using all the nodes in $S_h\setminus O_t(x_i)$. This is a standard problem which we solve using Dijkstra's algorithm \cite{dijkstra1959note}. 

\section{Empirical Study}

We test the validity of our peer--cohesion hypothesis on a number of stock indices and currency exchange rates. The objective is two--fold: first we aim at demonstrating that predictors perform better if they incorporate knowledge about the signals of their peers. The second is to devise a trading strategy based on the predictions. This serves as another way of demonstrating that the predictions contain extra information not available to individual nodes. 

Our procedure is to (a) define relevant markets (stock indices or currencies) and group them, (b) compute predictions of returns of group members and (c) take a (long--only) position if the signal node to which the member is connected shows a ``1''. Signal nodes consist mostly of macro--economic variables and will be introduced below. The situation we have in mind is that group members periodically leave and return to the group, forming a kind of fluctuation around equilibrium group membership levels. The fluctuation can be explained by the asynchronous way in which group members react to changes in the macro--economic environment. 

As an example, if the dominant narrative puts trade wars in the foreground, the automotive sector may be the first to price--in any progress or set--back in the matter while consumer stocks may follow. It should be noted that the role of leader and follower within a peer group does depend on the narrative: if trade wars turn into currency wars (resulting in a global pressure to keep rates down) then rate--sensitive sectors might be the ones leading market reactions. In general, the role of leader or follower changes over time and is not known a priori, even though, in some cases, regional leaders exist (US for developed markets, Brazil for LATAM etc.). 

As part of our future research program, we hope to isolate markets where a ground truth leader is known and can be used to test the precision of proposed scheme. In the experiments conducted so far, we did not notice any special role of any particular market. This may be attributed to the fact that, for now, the peer group definition is done in an ad--hoc fashion using traditional classifications such as defensive or cyclical stocks and developed vs. emerging (currency) markets. The experiments reported below show that in many cases, peers help detect important regime--changes in the market and cause the (long--only) strategy to stay out of the market during periods of crisis. An exhaustive analysis on the peer group definition and how it affects the benefits to its members is currently work in progress. 

We obtain data starting in Jan. 2002 from MSCI and Datastream. This allows us to measure a long--term signal co--dependence from which we build the signal tree $S_h$. The estimation is carried out in--sample (using all available data) which is still a weakness of the experiment although we argue that the estimation should be robust as the sample period represents more than a full economic cycle. We come back to this point in the section on further research at the end of the paper. Since the problem is one of signal selection we benchmark our results against two naive strategies. The first benchmark is to attach the target node to the signal node corresponding to the largest entry in the co--occurrence matrix without applying Dijkstra's algorithm (benchmark: \textit{greedy}). As a second reference, we include the performance of the underlying market (benchmark: \textit{underlying}), i.e. no attachment. The tree in this experiment is composed of the signals and asset returns listed below. Across the paper, signals are denoted by the prefix $\_$ while the target and the peers start with the marker $x$.

\subsection{Stocks: defensive and cyclical sectors}

We present results obtained for the defensive and cyclical sectors in the US and European equity markets. %
%The following abbreviations are used in the result charts, Fig. 1. \_AA\_XX: denotes the XX signal for the country AA. xAA\# denotes return of asset \# in country AA. The country abbreviations are CL-Chile, CO-Colombia, BR-Brazil, PR-Peru and MX-Mexico.  The XX abbreviations are described below. The non country specific signals are denoted by  \_YY where YY are described below. As all target variables belong to the same geographical region they are all included to belong to the same group $N(x)$. In our experiments we pick one of the nodes $x_i$ in $N(x)$ and treat all other nodes as its peers. 
%
The same set of signals is used in both studies.

\vspace{2ex}
\underline{Defensive and cyclical sectors: signals}
\vspace{1ex} 
\begin{enumerate}
\item \_RV.BD : Revenue indicator in Germany
\item \_RV.US : Revenue indicator in US
\item \_MO.BD : Momentum indicator in Germany
\item \_MO.US : Momentum indicator in US
\item \_EMO.BD : Earnings Momentum indicator in Germany
\item \_EMO.US : Earnings Momentum indicator in US
\item \_UN.US : Unemployment indicator in US
\item \_NO.US :  New orders indicator in US
\item \_ISM.BD : Manufacturing indicator in Germany
\item \_ISM.US : Manufacturing indicator in US
\item \_CF.BD : Confidence indicator in Germany
\item \_CF.US : Confidence indicator in US
\item \_STP.BD : Treasury steepness indicator in Germany
\item \_STP.US : Treasury steepness indicator in US

\end{enumerate}

All ``indicators'' correspond to 1Y z--scores of the underlying variable. The sign of the node \_UN.US is inverted to obtain a compatible set of indicators.

\subsubsection{Defensive sectors}
\begin{figure}[!t]
	\begin{center}
		\centerline{\includegraphics[width=1\columnwidth]{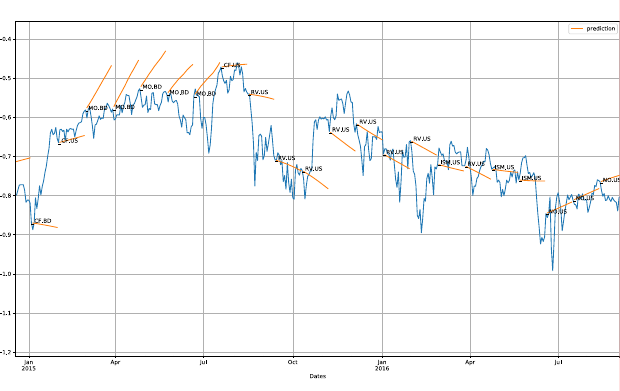}}
		\caption{Communication sector performance in Germany (node $xCO.BD$) with 20--days--ahead predictions based on the indicated signal node}.
		\label{tp2}
	\end{center}
\end{figure}

\begin{figure}[!t]
 	\begin{center}
 		\centerline{\includegraphics[width=.9\columnwidth]{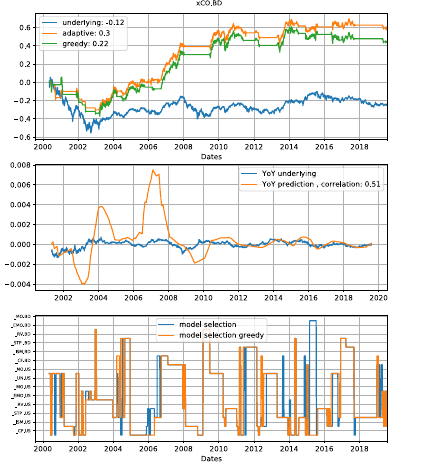}}
		\caption{Performance of the communication sector in Germany. The chart shows the performance of the proposed ``adaptive'' scheme as compared to the benchmarks (top), the year on year prediction compared to the underlying (middle) and the evolution of the time--varying attachment points of the strategy as compared to a greedy attachment}.
 		\label{tp1}
 	\end{center}
\end{figure}

\begin{figure}[!t]
 	\begin{center}
 		\centerline{\includegraphics[width=0.7\columnwidth]{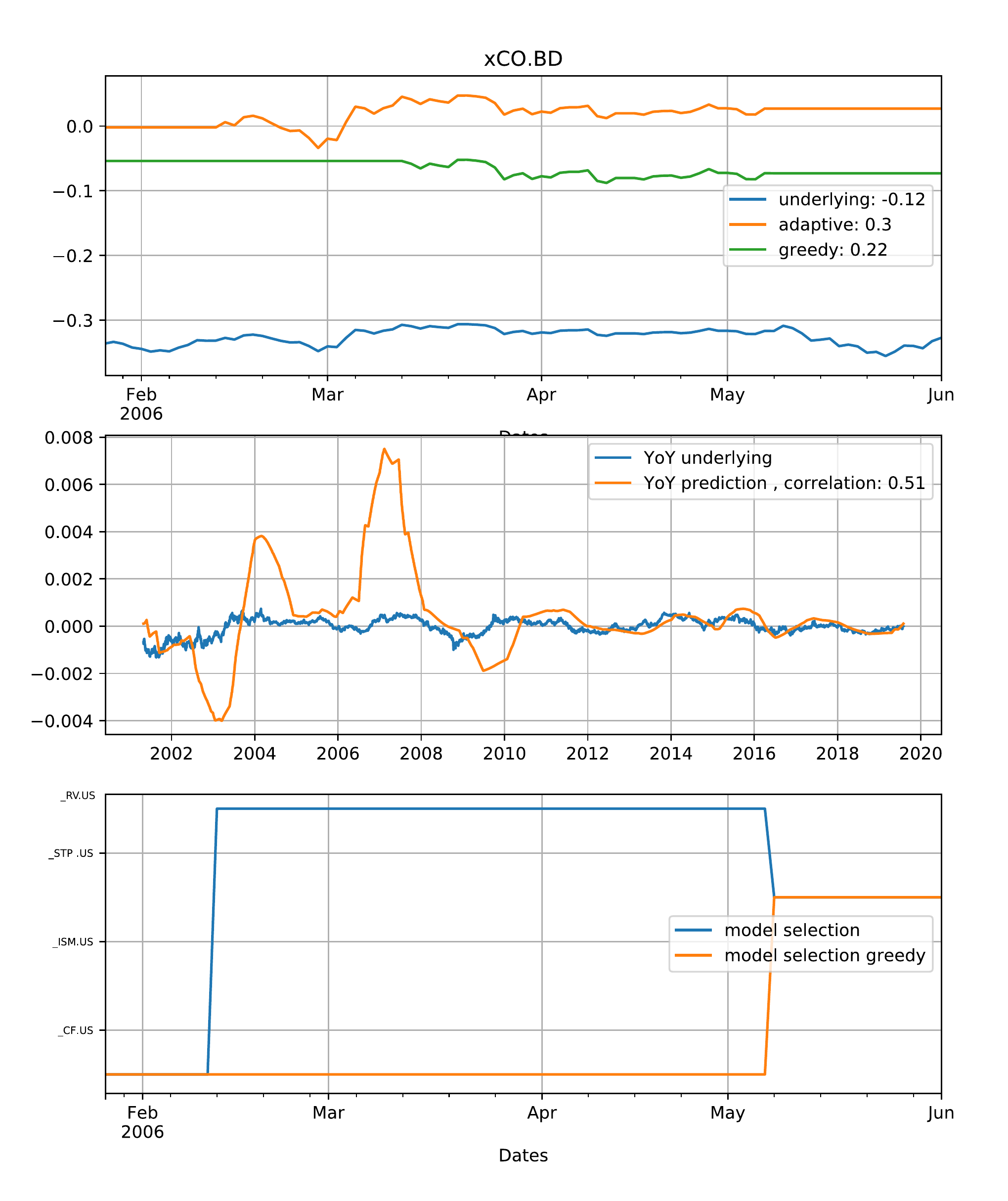}}
		\caption{Zoom on the Performance in 2006}.
 		\label{tp3}
 	\end{center}
\end{figure}

In this example, our target is the communication sector in Germany, denoted as $xCO.BD$. The set of peers includes all German and US defensive sectors and is defined as follows: 

\underline{Defensive peer group:}
\vspace{1ex} 
\begin{enumerate}
\item xUT.BD : Utilities sector in Germany
\item xUT.US : Utilities sector in US
\item xCO.BD : Communication sector in Germany
\item xCO.US : Communication sector in US
\item xHC.BD : Health care sector in Germany
\item xHC.US : Health care sector in US
\item xCS.BD : Consumer staples sector in Germany
\item xCS.US : Consumer staples sector in US
\end{enumerate}
The prediction resulting from peer--cohesion is presented in Figure \ref{tp2}. The cumulative return of our target is shown in blue. Every 20 days, we re--estimate the attachment of our target to the signal tree as described in section 3.3. For each period, we show the 20--days--ahead prediction in red as well as the name of the signal node (model) used for the prediction. The magnitude of the prediction is based on equalizing the (200 days trailing) volatility of the target and the signal. 
The investment strategy derived from the signals switches between a long--position in the underlying market or cash. At every instant, our target market connects to a node $s_j$ which is then evaluated. If $s_j=1$, a position is entered else we do not invest. While the signal tree is the same for every target market, the attachment points $s_j$ are different and depend on the cost for a target node to be close to its peers. We estimate the attachment of our target to the tree based on a rolling window of $200$ days in steps of $20$ days. Figure \ref{tp1} displays the cumulative profit and loss of the resulting adaptive investment strategy. The corresponding switching sequence is reported in the lower part of the chart. The middle chart compares the prediction against the actual, realized year--on--year return. The correlation turns out to be $0.51$. 
To further illustrate the mechanism of the proposed strategy, we provide a more detailed study of the example in Figure \ref{tp3}. We consider the period of Feb. 2006 where the adaptive strategy performs well while the greedy version looses.
In the lower chart of Figure \ref{tp3}, it can be seen that in Feb. 2006, both the greedy and the adaptive versions coincide and connect to the node \textit{\_CF.US}. In mid Feb. 2006, the adaptive version connects to \textit{\_RV.US} while the greedy version stays at \textit{\_CF.US}. The (minimum spanning) trees before and after the switch are seen in Figs.\ref{Ng1} and \ref{Ng2}: at the beginning of the month \textit{xCO.BD} connects to \textit{\_CF.US} which also corresponds to the simple shortest path (without awareness of the peer group). This is the situation in Fig.\ref{Ng1}. Later that month, we see (Fig.\ref{Ng2}) that a new shortest path connects our target to \textit{\_RV.US} while the greedy algorithm remains connected to \textit{\_CF.US} as shown by the dotted line. Note that this result is in--line with expectations since some peers also connect to \textit{\_RV.US} as an early sign of a new orientation of the peer group. The algorithm causes a re--wiring of the connection enabling the target node to include new information beyond \textit{\_CF.US} with which it continues to have the highest direct (greedy) co--occurrence. This is precisely the kind of group information transfer that we are exploring in this paper. The thickness of the edges represent the strength of the co--occurrences.  
\begin{figure*}[!t]
\centering
   \begin{subfigure}[b]{\textwidth}
 \centerline{\includegraphics[width=.65\columnwidth]{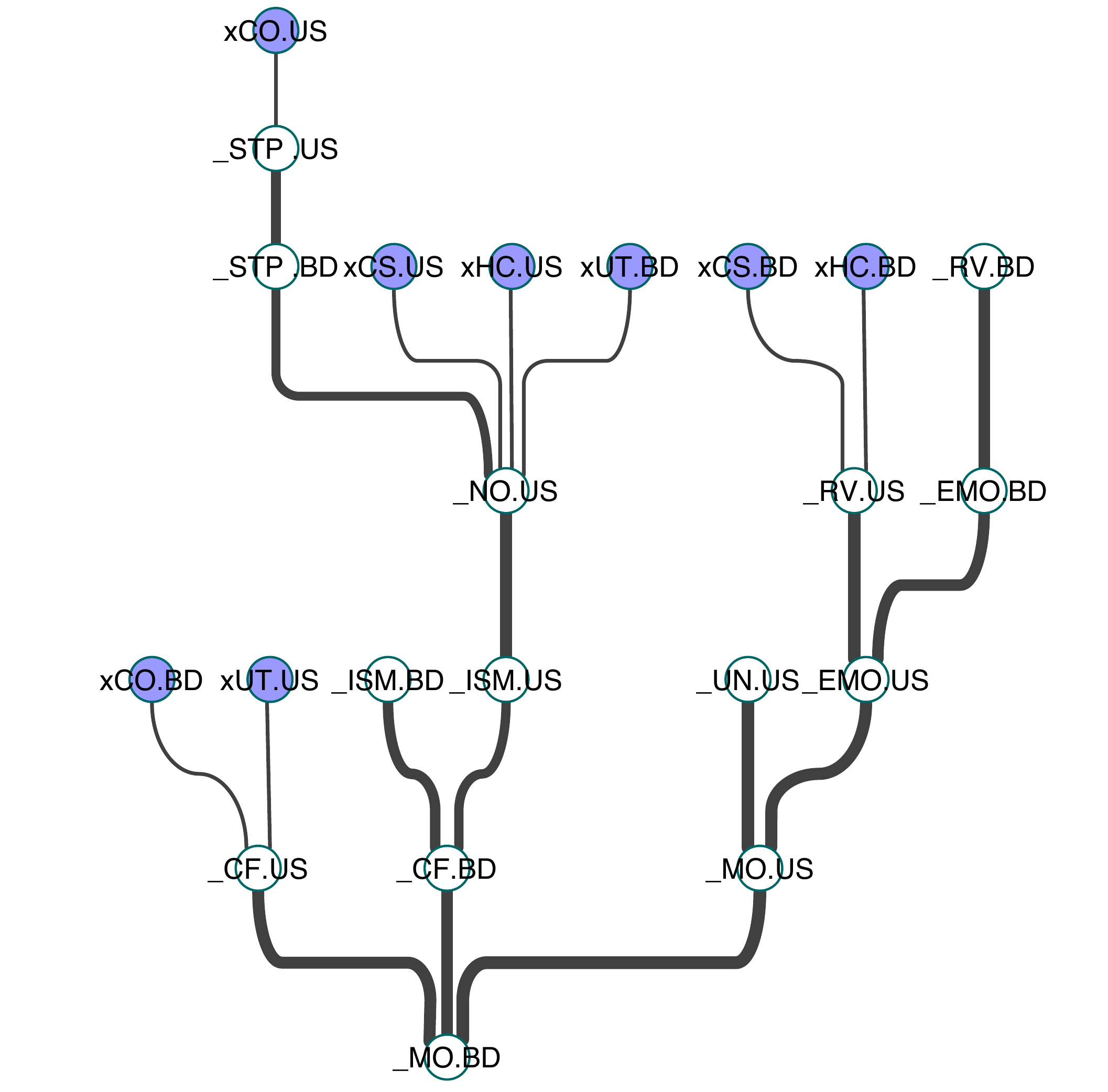}}
   \caption{}
   \label{Ng1} 
\end{subfigure}

\begin{subfigure}[b]{\textwidth}
\centerline{\includegraphics[width=.65\columnwidth]{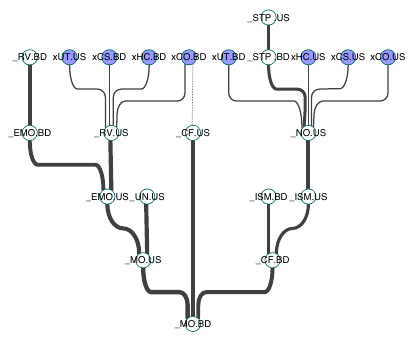}}
   \caption{}
   \label{Ng2}
\end{subfigure}
\caption{Minimum spanning trees estimated with target note attached during the period around Feb 2006. }
 \label{figd} 
\end{figure*}

\subsubsection{Cyclical sectors}
We now proceed to a similar study for the cyclical sectors. 

\vspace{2ex}
\underline{Cyclical peer group:}
\vspace{1ex} 
\begin{enumerate}
\item xIN.BD : Industrials sector in Germany
\item xIN.US : Industrials sector in US
\item xCD.BD : Consumer discretionary sector in Germany
\item xCD.US : Consumer discretionary sector in US
\item xMA.BD : Materials sector in Germany
\item xMA.US : Materials care sector in US
\item xIT.BD : Information technology  sector in Germany
\item xIT.US : Information technology sector in US
\end{enumerate}

\begin{figure}[!ht]
	\begin{center}
		\centerline{\includegraphics[width=1\columnwidth]{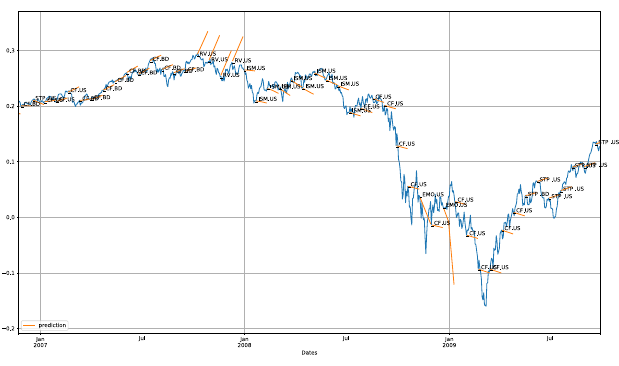}}
		\caption{Industrials performance in the U.S. (node $xIN.US$) with 20--days--ahead predictions based on the indicated signal node}.
		\label{tp12}
	\end{center}
\end{figure}

\begin{figure}[!t]
 	\begin{center}
 		\centerline{\includegraphics[width=.9\columnwidth]{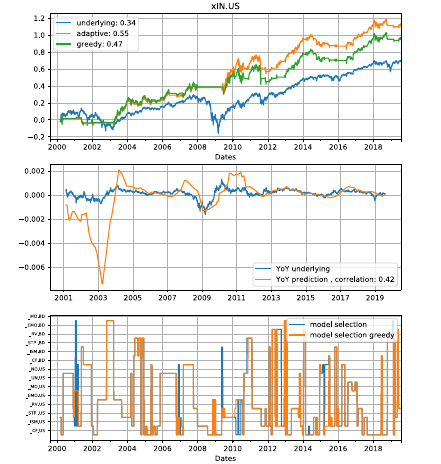}}
		\caption{Performance of the industrial sector in the US. The chart shows the performance of the proposed ``adaptive'' scheme as compared to the benchmarks (top), the year--on--year prediction compared to the underlying (middle) and the evolution of the time--varying attachment points of the strategy as compared to a greedy attachment}.
 		\label{tp11}
 	\end{center}
\end{figure}

Our target is the industrial sector in the US \textit{xIN.US}. The year on year prediction in Fig.\ref{tp11} presents a correlation of $.42$ with the underlying. Fig.\ref{tp12} provides a detailed view of the 20--day predictions and the name of the forecasting model from which they are derived. 

\begin{figure}[h]
 	\begin{center}
 		\centerline{\includegraphics[width=.7\columnwidth]{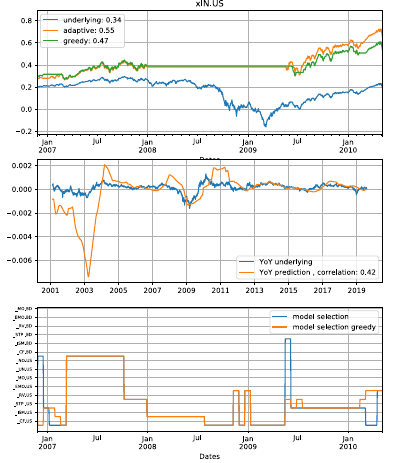}}
		\caption{Zoom on the Performance in 2008}.
 		\label{tp13}
 	\end{center}
\end{figure}

\begin{figure}[h]
 	\begin{center}
 		\centerline{\includegraphics[width=.7\columnwidth]{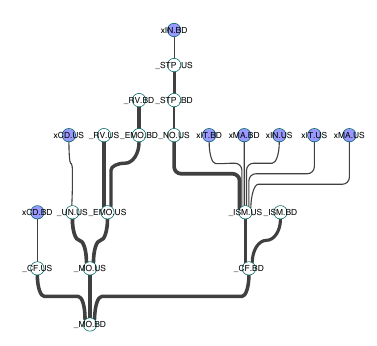}}
		\caption{\textit{xIN.US} tree at the beginning of 2008}.
 		\label{tp14}
 	\end{center}
\end{figure}

Fig.\ref{tp13} provides some instructive details on the strategy performance during the great financial crisis.
The adaptive scheme stays in cash because the signal coming from \textit{\_ISM.US} turns negative. The tree corresponding to this period is shown in Fig. \ref{tp14}.
Later in 2008, our target connects to \textit{\_CF.US} which also causes the strategy to stay in cash and further out--perform the outright long strategy. In the following year, the adaptive strategy manages to catch the rebound by connecting to \textit{\_STP.BD} and subsequently to \textit{\_RV.US} which both turn positive. 
 
The results provide some support for the adaptive (``learning from others'') approach but are by no means exhaustive. It seems that following peers adds value when choosing a forecasting model. The robustness of the result and its dependence on critical parameters, such as the choice of the peer group, the lookback horizons used for the indicators and the estimation of co--occurrences need to be investigated further. As mentioned above, this is currently work in progress.

\begin{figure*}[!thb]
	\centering
	\begin{subfigure}[b]{\textwidth}
		\centerline{\includegraphics[width=.85\columnwidth]{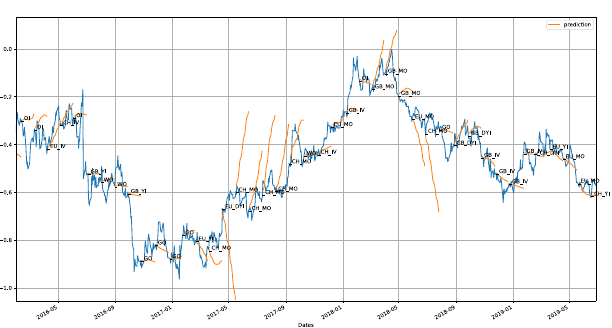}}
		\caption{}
		\label{Ng1} 
	\end{subfigure}
	
	\begin{subfigure}[b]{\textwidth}
		\centerline{\includegraphics[width=.85\columnwidth]{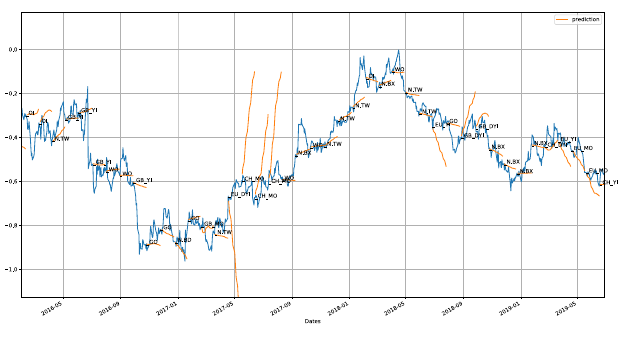}}
		\caption{}
		\label{Ng2}
	\end{subfigure}
	\caption{Cumulative returns and predictions over time (a) without news (b) with news signals. }
	\label{figd2} 
\end{figure*}

\subsection{Currencies: GBP vs. USD}
As an illustration of the universality of the approach, we study a similar prediction task for the \textit{GBPUSD} currency pair. The peer group in this case is composed of some important European countries: 

\vspace{2ex}
\underline{GBP peer group:}
\vspace{1ex} 
\begin{enumerate}
	\item xGB2 : British Pound against the USD spot return
	\item xEU2 : Euro against the USD spot return
	\item xCH2 : Swiss Franc against the USD spot return
\end{enumerate}

The set of signals consists of index returns and news data obtained from our proprietary news database \footnote{Daily data obtained from $\sim$100 pre--defined RSS feeds since March 2015.}. The news signals count keywords in a corpus of financial blogs and news articles. 

\begin{enumerate}
\item \_WO : MSCI World price return
\item \_CL : Crude Oil price return
\item \_GC : Gold price return
\item \_EM : MSCI emerging markets price return
\item \_N.TW: News--based ``trade war'' indicator
\item \_N.BX: News--based ``brexit'' indicator
\item \_N.BD: News--based ``dovish (referring to monetary policy)'' indicator
\end{enumerate}
 
In addition we include some country specific indicators as follows:
\begin{enumerate}

\item YI (Carry): the difference between the money market deposit rate of a country vs. the US deposit rate

\item DYI (Change of Carry): the difference between the YI signal now and 60 days ago

\item MO (Momentum): average spot return of a currency where the indicator is 1 if $\langle r_{spot}\rangle_{100} > 0$ and $\langle\cdot\rangle_{100}$ is the arithmetic average over 100 days

\item IV (Implied Vol): implied volatility of a currency. The higher the implied vol, the more volatility is expected by market participants trading in options. The signal is obtained by applying a 1Y z--score on IV and invert the sign

\end{enumerate}

We compare the results with and without news in Fig. \ref{Ng1} and \ref{Ng2}. It can be seen that the news signals, if available, are quite often selected as attachment nodes. 

\section{Further directions}

Apart from investigating peer cohesion effects across different markets and time--scales, the authors believe that the assumption on the stationarity of $S_h$ can be relaxed. This would also be of practical relevance as $S_h$ could be estimated on trailing data, possibly on an expanding window as more data becomes available. The associated (slow) time-variation in $S_h$ would give rise to very interesting situations where peer groups may entirely be torn apart due to a change in the co--dependence structure of signal nodes that connected them.

\clearpage

\bibliographystyle{IEEEtran}
\bibliography{mybibfile22}
\end{document}